\documentclass[twocolumn,twocolappendix]{aastex631}
\usepackage{graphicx}
\usepackage{amsmath}
\usepackage{gensymb}
\usepackage{xspace}
\usepackage{hyperref}

\defcitealias{sgraI}{\sgra Paper I}
\defcitealias{sgraII}{\sgra Paper II}
\defcitealias{sgraIII}{\sgra Paper III}
\defcitealias{sgraIV}{\sgra Paper IV}
\defcitealias{sgraV}{\sgra Paper V}
\defcitealias{sgraVI}{\sgra Paper VI}
\defcitealias{sgraVII}{\sgra Paper VII}
\defcitealias{sgraVIII}{\sgra Paper VIII}

\def\lsim{\mathrel{\raise.3ex\hbox{$<$\kern-.75em\lower1ex\hbox{$\sim$}}}}
\def\gsim{\mathrel{\raise.3ex\hbox{$>$\kern-.75em\lower1ex\hbox{$\sim$}}}}

\def\m87{M87$^*$\xspace} 
\def\sgra{Sgr~A$^*$\xspace}

\begin{document}

\title{Prospects of Detecting a Jet in Sagittarius A* with VLBI}

\author{Erandi Chavez}
\affiliation{Center for Astrophysics $|$ Harvard \& Smithsonian, 60 Garden St, Cambridge, MA 02138, USA}
\author{Sara Issaoun}
\affiliation{Center for Astrophysics $|$ Harvard \& Smithsonian, 60 Garden St, Cambridge, MA 02138, USA}
\affiliation{NASA Hubble Fellowship Program, Einstein Fellow}
\author{Michael D. Johnson}
\affiliation{Center for Astrophysics $|$ Harvard \& Smithsonian, 60 Garden St, Cambridge, MA 02138, USA}
\author{Paul Tiede}
\affiliation{Center for Astrophysics $|$ Harvard \& Smithsonian, 60 Garden St, Cambridge, MA 02138, USA}
\author{Christian Fromm}
\affiliation{Institut f\"ur Theoretische Physik und Astrophysik, Universit\"at W\"urzburg, Emil-Fischer-Strasse 31, 97074
W\"urzburg, Germany}
\affiliation{Institut f\"ur Theoretische Physik, Goethe Universit\"at, Max-von-Laue-Str. 1, D-60438 Frankfurt, Germany }
\affiliation{Max-Planck-Institut f\"ur Radioastronomie, Auf dem H\"ugel 69, D-53121 Bonn, Germany}
\author{Yosuke Mizuno}
\affiliation{Tsung-Dao Lee Institute, Shanghai Jiao-Tong University, Shanghai, 520 Shengrong Road, 201210, People's Republic of China}
\affiliation{School of Physics \& Astronomy, Shanghai Jiao-Tong University, Shanghai, 800 Dongchuan Road, 200240, People's Republic of China}
\affiliation{Institut f\"ur Theoretische Physik, Goethe Universit\"at, Max-von-Laue-Str. 1, D-60438 Frankfurt, Germany }

\begin{abstract}
Event Horizon Telescope (EHT) images of the horizon-scale emission around the Galactic Center supermassive black hole Sagittarius~A* (\sgra) favor accretion flow models with a jet component. However, this jet has not been conclusively detected. Using the ``best-bet'' models of \sgra from the EHT collaboration, we assess whether this non-detection is expected for current facilities and explore the prospects of detecting a jet with VLBI at four frequencies: 86, 115, 230, and 345\,GHz. We produce synthetic image reconstructions for current and next-generation VLBI arrays at these frequencies that include the effects of interstellar scattering, optical depth, and time variability. We find that no existing VLBI arrays are expected to detect the jet in these best-bet models, consistent with observations to-date. We show that next-generation VLBI arrays at 86 and 115\,GHz---in particular, the EHT after upgrades through the ngEHT program and the ngVLA---successfully capture the jet in our tests due to improvements in instrument sensitivity and ($u,v$) coverage at spatial scales critical to jet detection. These results highlight the potential of enhanced VLBI capabilities in the coming decade to reveal the crucial properties of \sgra and its interaction with the Galactic Center environment. 

\end{abstract}


\section{Introduction} \label{sec:intro}

Sagittarius~A* (\sgra) was first discovered in 1974 as a compact radio source in the Galactic Center \citep{balickandbrown1974}. Decades of infrared observations tracking stellar orbits in the innermost regions about \sgra have shown that it is an extremely compact and massive object, consistent with being a supermassive black hole (SMBH) with mass $M \sim 4 \times 10^{6}\,M_\odot$ \citep{schodel2002,ghez2003,ghez2008,Do2019,Gravity_2022}. Since its discovery, multi-wavelength observations ranging from radio to X-ray (excluding optical and UV due to severe attenuation by dust) have measured the spectrum of \sgra with the goal of constraining its intrinsic properties \citep[see, e.g.,][]{sgraII}. The shape of the radio spectrum in particular is indicative of partially self-absorbed synchrotron radiation, where an excess of emission at submillimeter wavelengths, called the ``submillimeter bump,'' corresponds to the compact region closest to the black hole \citep[e.g.,][]{Zylka_1992,Krichbaum_1998,Falcke1998, bower2006, doeleman2008, sgraI}. 

Two classes of models describing the source emission of \sgra at near-horizon scales successfully reproduce the basic properties inferred from its spectrum. The first describes \sgra as a compact, relativistic jet \citep{BlandfordandKonigl1979,falckeandmarkoff2000} while the second describes an advection dominated accretion flow \citep[ADAF;][]{narayan1995sgra,Yuan_2003}. In the jet models, emission originates primarily from the jetted outflow, while in the ADAF models the emission originates from a hot, geometrically thick accretion disk. To break the degeneracy between these models, detailed information about the source morphology at near-horizon scales is needed. The close proximity of \sgra permits direct imaging of its innermost accretion structure using Very Long Baseline Interferometry (VLBI) at millimeter-wavelengths, which presents an opportunity to clarify the true nature of the inner accretion flow of \sgra. 

Recent millimeter VLBI and multi-wavelength efforts using observations from the 2017 Event Horizon Telescope (EHT) campaign have now imaged horizon-scale structure of \sgra in total intensity and polarization \citep[hereafter known as \sgra Papers I-VIII]{sgraI, sgraII, sgraIII, sgraIV, sgraV, sgraVI, sgraVII, sgraVIII}. The interpretation of EHT images has heavily relied on general relativistic magnetohydrodynamic (GRMHD) simulations, which, unlike the models mentioned above, self-consistently include both disk and jet components \citep{Gammie_2003,Porth_2019}. In the exploration of a large parameter space of models, two general solutions emerge: one in which the radio emission is primarily generated by the accretion disk (near the equatorial plane), and another in which emission is primarily generated by the jet (close to the system spin axis). The best-bet models resulting from this exploration successfully reproduce most properties of \sgra, including the image morphology found by the EHT, and suggest that the emission at radio wavelengths originates predominantly from the accreting material near the black hole \citepalias[see][]{sgraV}. These same simulations predict the presence of a faint, extended jet. Observationally, there is speculative support for a jet from \sgra from much larger scale features, including some linear X-ray features on parsec scales \citep[e.g.,][]{Li_2013} and the ``Fermi bubbles'' seen in gamma rays on kiloparsec scales \citep{Su_2010}. However, a jet has yet to be conclusively detected with observations.

In this paper, we explore the possibility of directly imaging a jet in \sgra using VLBI. There are many technical and astrophysical challenges facing this proposition that we aim to characterize. The first is interstellar scattering. Ionized plasma between Earth and the Galactic Center acts as a scattering screen at radio wavelengths, which in turn obscures \sgra's intrinsic structure. The size of the scattering kernel is inversely proportional with wavelength, $\theta_\textrm{scatt} \sim (\textrm{1\,mas}) \times \lambda_\textrm{cm}^2$ \citep{Davies_1976,vanLangevelde_1992,Bower_2004,Shen_2005,bower2006,johnson2018}. For observations at wavelengths $\geq$3.5\,mm (86\,GHz), many studies over the past two decades have demonstrated that the observed structure is completely dominated by scattering effects \citep{Doeleman_2001,Shen_2005,Lu_2011,Ortiz_2016,Brinkerink_2016,johnson2018,issaoun2019}. Previously, \citet{Markoff_2007} has argued that non-detections of the jet in \sgra using VLBI at 43\,GHz can be attributed to the effects of scattering.

The second challenge is the intrinsic variability of \sgra. The period of \sgra's innermost stable circular orbit (ISCO) ranges from 4-53 minutes, which suggests that compact emission close to the black hole will be variable on similar timescales. This is supported by multi-wavelength observations. Light curves at near-infrared, far-infrared, and millimeter wavelengths show variability on intrahour-timescales \citep[e.g,][]{gravity2020,stone2016,wielgus2022}. Flaring events are also common, with $\sim$one flare per day occurring in X-rays and a few flares per day in the near-infrared \citep{neilsen2013}. Therefore observations taken over many hours, such as those by the EHT, will be significantly impacted by its intrinsic variability \citepalias{sgraI,sgraII,sgraIII,sgraIV}.

The third challenge is due to limitations in the capabilities of existing VLBI arrays. Instrumental effects, such as sensitivity and imperfect baseline coverage, all significantly impact the possibility of jet detection. While VLBI arrays have not yet detected a jet in \sgra, proposed next-generation VLBI arrays such as the EHT after upgrades through the ngEHT program\footnote{For simplicity, we will refer to the upgraded EHT as the ngEHT throughout this paper and will use EHT to refer to the EHT array as of 2023.} and the ngVLA will provide improvements in sensitivity and $(u,v)$ coverage that could enable the first jet detections \citep[e.g.,][]{selina2018,doeleman2023,johnson2023ngeht}. 

In this paper, we explore the ability to detect and image a jet as predicted by the best-bet GRMHD model from \citetalias{sgraV}, taking into account interstellar scattering, intrinsic variability, and array configurations. In Section~\ref{sec:grmhd}, we discuss the jet properties of the best-bet GRMHD model at four different frequencies (86, 115, 230 and 345\,GHz), along with how scattering and instrument resolution impact the prospects of jet detection. In Section~\ref{sec:vlbi_imaging}, we perform synthetic image reconstructions of \sgra with the Julia-based VLBI imaging software \texttt{Comrade} \citep{tiede2022} considering various frequency and array combinations to determine the best prospects towards jet detection. We present our imaging results, discussion, and overall conclusions in Sections~\ref{sec:results}, \ref{sec:discussion}, and \ref{sec:conclusions}, respectively.

\begin{figure*}
    \centering
    \includegraphics[width=1.0\textwidth]{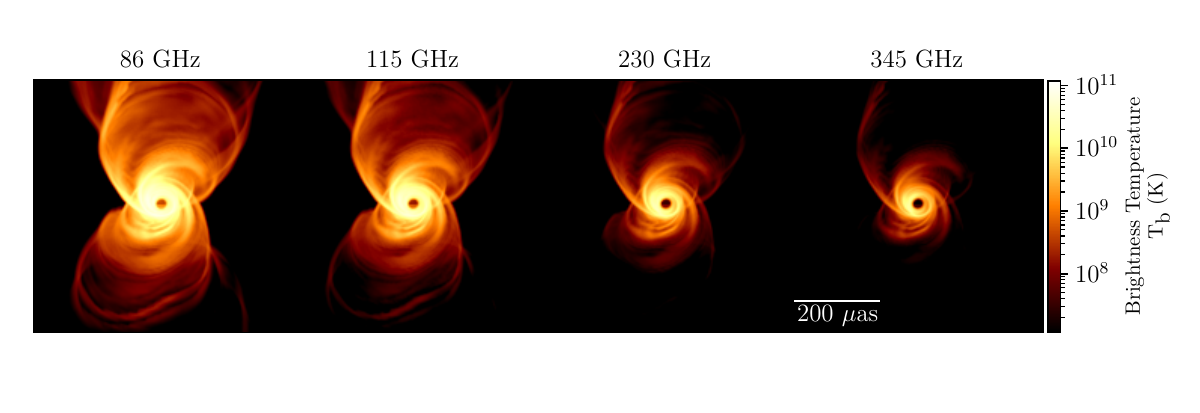}
    \caption{A single snapshot in time of the best-bet GRMHD simulation of \sgra, ray-traced at four frequencies: 86, 115, 230, and 345\,GHz. The parameters describing this model were taken from the ``best-bet'' parameters in \citetalias{sgraV}: a black hole spin of $a_\star = 0.94$, $R_\text{low} = 1$, $R_\text{high} = 160$, and an inclination of $30 \degree$. The images are shown with a field of view of $600 \times 600$ $\mu$as, with a logarithmic color scale to display the faint extended emission.}
    \label{fig:grmhdsnaphots}
\end{figure*}

\begin{figure*}
\includegraphics[width=1.0\textwidth]{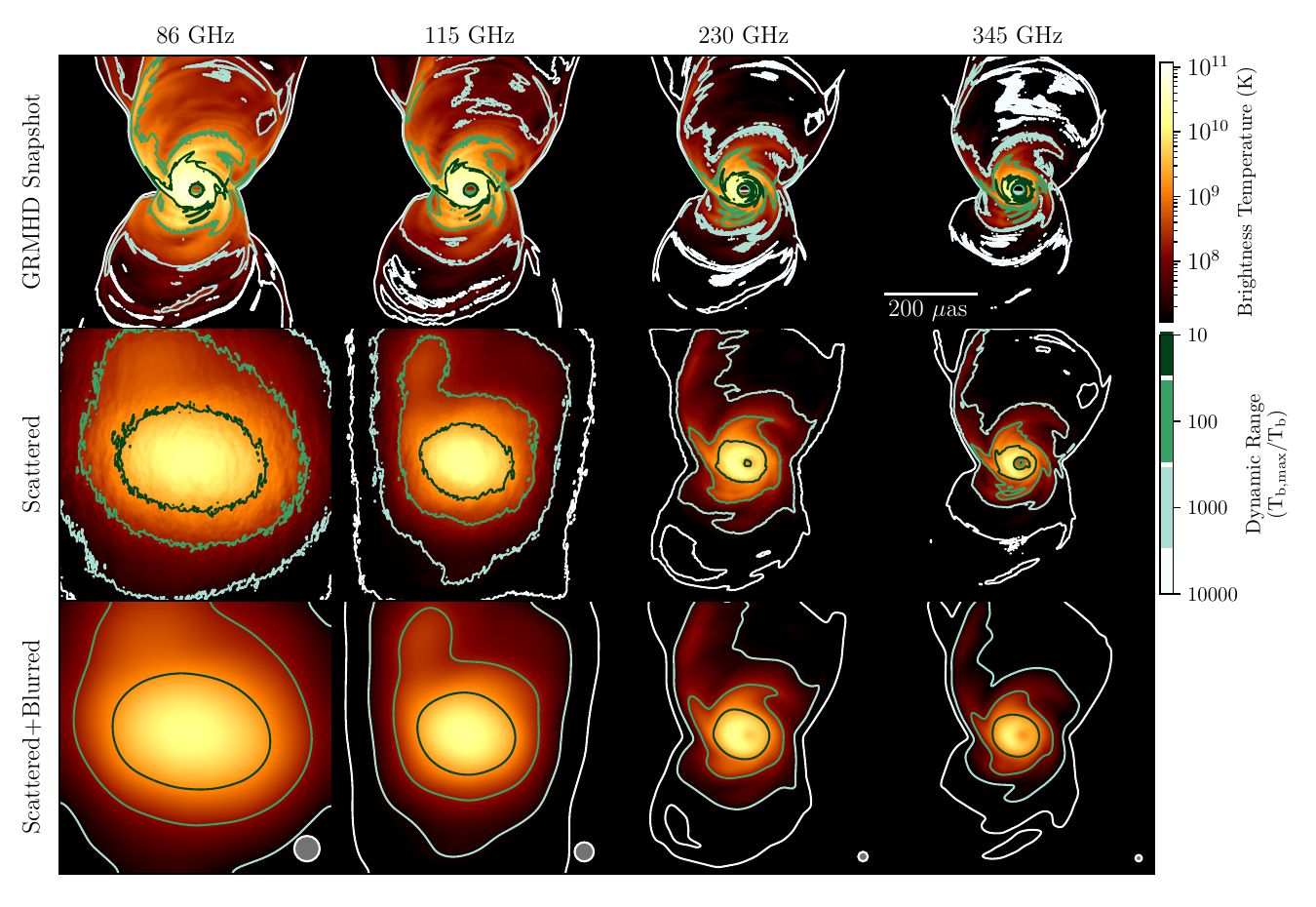}
\caption{A grid of snapshots of the best-bet model for \sgra described in Section \ref{sec:best-bet}. Top row: simulations at 86, 115, 230, and 345\,GHz. Middle row: The GRMHD snapshots, now convolved with the interstellar scattering screen \citep{johnson2016} to predict the on-sky appearance of \sgra.
Bottom row: The scattered GRMHD snapshots now blurred with circular Gaussian beams of diameter 56, 42, 21, and 12\,$\mu$as --- the resolution of an Earth-sized telescope at each frequency. This represents the theoretical limit of what an earth-sized telescope could observe at each frequency.}
\label{fig:grmhd_snapshot_contours_grid}
\end{figure*}

\section{Predicting the Appearance of a Jet in \sgra \label{sec:grmhd}}

\subsection{Intrinsic Source Model \label{sec:best-bet}}

The EHT Collaboration used a combination of EHT observations at 230\,GHz, GMVA observations at 86\,GHz, Very Large Telescope near-infrared observations at 2.2\,$\mu$m, and X-ray observations from the Chandra X-ray Observatory to constrain a wide variety of parameters describing images produced from ray-traced GRMHD models \citepalias{sgraI,sgraII,sgraV}. The ``best-bet'' parameters described in \citetalias{sgraV}, which passed the most constraints, are: prograde spin (dimensionless spins of $a_\star=0.5$ and $a_\star=0.94$ both pass), a strong magnetic field, an observer inclined $~30\degree$ from the black hole spin axis, and a plasma heating parameter $R_\textrm{high} = 160$ (see below). In this paper, we use these parameters and select the $a_\star=0.94$ simulation. Note that the position angle of the spin axis is unconstrained, corresponding to an arbitrary rotation of the intrinsic on-sky image.

Here we provide additional details on this model and its associated parameters. We used the state-of-the-art GRMHD code \texttt{BHAC} \citep{Porth2017} to simulate a magnetized fluid accreting onto a black hole with spin $a_\star=0.94$ in three dimensions \citep[for a comparison of GRMHD codes, see][]{Porth_2019}. \texttt{BHAC} solves the GRMHD equations using geometric units i.e., $G=M=c=1$. The simulation is performed in logarithmic Kerr-Schild coordinates $(\ln r, \theta, \phi)$ where the outer edge of the simulation box is located at $r=2500\,r_g$ where $r_g = GM/c^2$ is the Schwarzschild Radius. We initialize the simulation with a torus in hydrodynamic equilibrium placed in the equatorial plane of the black hole, where the inner edge of the torus is at $r=20\,M$ and the pressure maximum of the torus is located at $r=40\,M$ \citep{Fishbone76}. We initialize the torus with a weak, poloidal magnetic field \citep[see][for additional details]{Fromm2022}.

This torus setup leads to a magnetically arrested disk (MAD) state \citep{Narayan_2003,Igumenshchev_2003,Tchekhovskoy2012}. In order to resolve the magneto-rotational instability triggering the accretion, i.e., the angular momentum transport, we use three levels of adaptive mesh refinement, which translates to a resolution of $384\times192\times192$ in the radial, polar, and azimuthal direction. In order to secure a quasi-steady accretion rate we evolve the simulation to $30{,}000\,M$ \citep[for more detail, see][]{Fromm2022}. 

Once the quasi-steady state is obtained we perform the radiative transfer calculation in order to obtain the emission of the accreting black hole at various frequencies. For the radiative transfer calculation we use the code \texttt{BHOSS} \citep{Younsi2012,Younsi2020,Gold_2020}. Since the GRMHD simulations are scale-free (see above), we scale our simulation to the \sgra using the black hole mass, $M=4.14\times10^6\,M_\odot$ and its distance $D=8.127\,\mathrm{kpc}$ and set the observers viewing inclination of 30$^\circ$ relative to the black hole spin axis. In the next step, we assume a thermal electron distribution function and compute the fractional temperature of the radiating electrons relative to the protons using a simple phenomenological formula that varies with the local gas-to-magnetic pressure ratio \citep{Moscibrodzka2016}. This formula is determined by two parameters: $R_\mathrm{low}$ and $R_\mathrm{high}$, which give the proton-to-electron temperature ratios in the strongly and weakly magnetized limits, respectively. The best-bet model has $R_\mathrm{low}=1$ and $R_\mathrm{high}=160$. 

In order to avoid contamination of our radiative transfer calculation by numerical floors in the jet spine, we exclude emission from regions where the magnetization $\sigma>1$ ($\sigma \equiv b^2/\rho$, where $b$ is the magnetic field energy density, and $\rho$ is the plasma rest mass energy density). In contrast to \citetalias{sgraV}, we increase the field of view (FOV) to 600\,$\mu$as to capture extended jet emission. Finally, we solve for the mass accretion rate that gives an average flux density of 2.4\,Jy at 230\,GHz (\citealt{wielgus2022}; \citetalias{sgraV}). Using the resulting accretion rate, $\dot{M}=2.54\times10^{-9}\,M_\odot$, we generate movies at four frequencies: 86, 115, 230, and 345\,GHz. The first three correspond to common radio observing bands, while 345 GHz is an expected expansion of the EHT array \citep{doeleman2023,johnson2023ngeht}.

The total simulation length is $1000M$ (where $M = GM/c^3 = r_g/c$, this corresponds to 5 hours and 40 minutes for the mass of \sgra), with each snapshot separated by $10M$ (204 seconds). \autoref{fig:grmhdsnaphots} shows an example of snapshot images from the simulation at each of the four frequencies. These frequencies were chosen to determine jet detection prospects with multiple existing and proposed VLBI arrays. We perform the majority of our analyses in the idealized scenario of a static source, i.e., reconstructing the snapshot images shown in \autoref{fig:grmhdsnaphots}. We discuss the additional challenges of source variability on jet detection in \autoref{sec:discussion}.

\subsection{On-Sky Image \label{sec:onskyimage_scattering}}

While the GRMHD simulation predicts the intrinsic structure of \sgra, an accurate prediction of the source image on the sky also requires simulating the effects of interstellar scattering toward the Galactic Center. Density inhomogeneities and turbulence in the ionized interstellar medium change the phase of incoming radio waves originating from the Galactic center, which results in a scattered image \citep[see e.g.,][]{Rickett_1990,Narayan_1992,thompson2017interferometry}. This scattering has two distinct effects on observed source structure: the first is large-scale, anisotropic blurring from small-scale density fluctuations ($\ll 0.1\,{\rm AU}$) dubbed ``diffractive scattering'', and the second is the introduction of small-scale image distortions and substructure from larger-scale density fluctuations ($\gsim 0.1\,{\rm AU}$) dubbed ``refractive scattering" \citep[see, e.g.,][]{NarayanGoodman89, GoodmanNarayan89,Johnson_2015,Psaltis_2018,johnson2018}.

The blurring due to diffractive scattering has been measured at radio wavelengths for decades \citep[e.g.,][]{Davies_1976,vanLangevelde_1992,Frail_1994,Bower_2004,Shen_2005,Bower_2006,johnson2018}. This diffractive effect is approximated by an elliptical Gaussian kernel with a full width at half maximum (FWHM) of $\theta_{\rm scatt} = (1.380 \pm 0.013) \lambda_\text{cm}^2$ along the major axis, a FWHM of $\theta_{\rm scatt} = (0.703 \pm 0.013) \lambda_\text{cm}^2$ along the minor axis, and a major axis position angle (PA) of $81.9 \degree$ east of north, resulting in more severe blurring in the east-west direction within the image \citep{johnson2018}. Refractive scattering introduces a small-scale substructure in the image, which results in a ``crinkling'' or ``wrinkling'' effect. The $\lambda^2$ scaling of the scattering kernel results in more severe scattering at lower observing frequencies; at frequencies lower than 86\,GHz, the intrinsic structure of \sgra is sub-dominant to scattering. The middle row of Figure~\ref{fig:grmhd_snapshot_contours_grid} displays the impact of scattering on a GRMHD snapshot at the four observing frequencies we investigate.

We use \texttt{Stochastic Optics} \citep{johnson2016} to model the effects of scattering on our image structure. \texttt{Stochastic Optics} is a module in the Python package \texttt{eht-imaging} \citep{chael2018} that implements the scattering model for \sgra \citet{Psaltis_2018,johnson2018}. The effects of scattering on the GRMHD snapshot at the four observing frequencies we investigate are shown in the middle row of Figure~\ref{fig:grmhd_snapshot_contours_grid}. As expected, the $\lambda^2$ scaling of the scattering kernel means that the lower frequencies are significantly more obscured by interstellar scattering. While the emission from the jet is significantly stronger at the lower frequencies, the diffractive blurring causes the emission to appear more smeared. Despite this effect, extended emission from the jet still persists at all frequencies at dynamic ranges of $\sim100$.

We also simulate the effects of instrument resolution on the images by blurring each snapshot with a circular Gaussian beam, where the FWHM corresponds to the resolution of an Earth-sized array at that frequency. The FWHM of these Gaussian beams correspond to 56, 42, 21, and 12\,$\mu$as at 86, 115, 230, and 345\,GHz, respectively (see bottom row of Figure~\ref{fig:grmhd_snapshot_contours_grid}). The resulting scattered and blurred images approximate what a perfect (diffraction-limited)) earth-sized telescope would observe at each frequency, and represent the idealized limit of recoverable jet structure through imaging \sgra. In practice, high-frequency VLBI arrays only sample a sparse combination of spatial frequencies on the sky, which needs to be taken into account for realistic expectations of jet recovery.

\section{Assessing the Prospects of VLBI Imaging \label{sec:vlbi_imaging}}

Interferometric measurements are made simultaneously with pairs of stations in an array. Each pair of stations forms a single baseline $\textbf{u} = \textbf{d} / \lambda$, where $\textbf{d}$ is the distance vector of length $d$ between the two stations, projected onto the plane that lies orthogonal to the line of sight, and $\lambda$ is the observing wavelength. The baseline vector $(\textbf{u})$ can be decomposed into an east-west $(u)$ and a north-south $(v)$ component.

Each baseline $(u,v)$ in an ideal interferometer measures the complex visibility $\mathcal{V}(u,v)$:
\begin{equation}
    \mathcal{V}(u,v) = \iint \, I(x,y) \text{e}^{-2 \pi i (ux+vy)} dx dy
    \label{eq:idealcomplexvis}
\end{equation}
where $I(x,y)$ is the on-sky image at the observing frequency and $(x,y)$ are the angular sky coordinates measured in radians \citep[e.g., ][]{thompson2017interferometry}. Hence, an interferometer samples Fourier components of the on-sky image. As a result, each baseline is sensitive to on-sky structure on angular scales of $1/u = \lambda/d$; VLBI arrays can thereby sample fine angular structures using extremely long baselines. 

For an ideal interferometer with perfect sampling, the on-sky image can be obtained from the complex visibilities by inverting \autoref{eq:idealcomplexvis}. However, in VLBI many spatial scales remain unsampled due to the limited number of stations in an array, which in turn makes the retrieval of $I(x,y)$ challenging. In addition, the complex visibilities measured by real interferometers are corrupted by thermal noise and systematic effects local to each station. With these corruptions, the complex visibility measured by a baseline formed by stations $i$ and $j$ can be written as:
\begin{equation}
    V_{ij} \approx g_i g_j^* \, \mathcal{V}_{ij} + \sigma_{ij}
    \label{eq:measuredcomplexvis}
\end{equation}
where $\mathcal{V}_{ij} = \mathcal{V}(u_{ij},v_{ij})$ is the source visibility on that baseline, $\sigma_{ij}$ is thermal noise associated with the measurement, and $g_i$, $g_j$ are complex numbers (called ``gains''), which encompass other forms of instrumental corruptions present in the data \citep[e.g., ][]{thompson2017interferometry}. 
Hence, both the array configuration and the individual telescope sensitivities affect what intrinsic source structure in the on-sky image can be recovered via imaging.

\begin{figure*}
    \centering
    \includegraphics[width=1\textwidth]{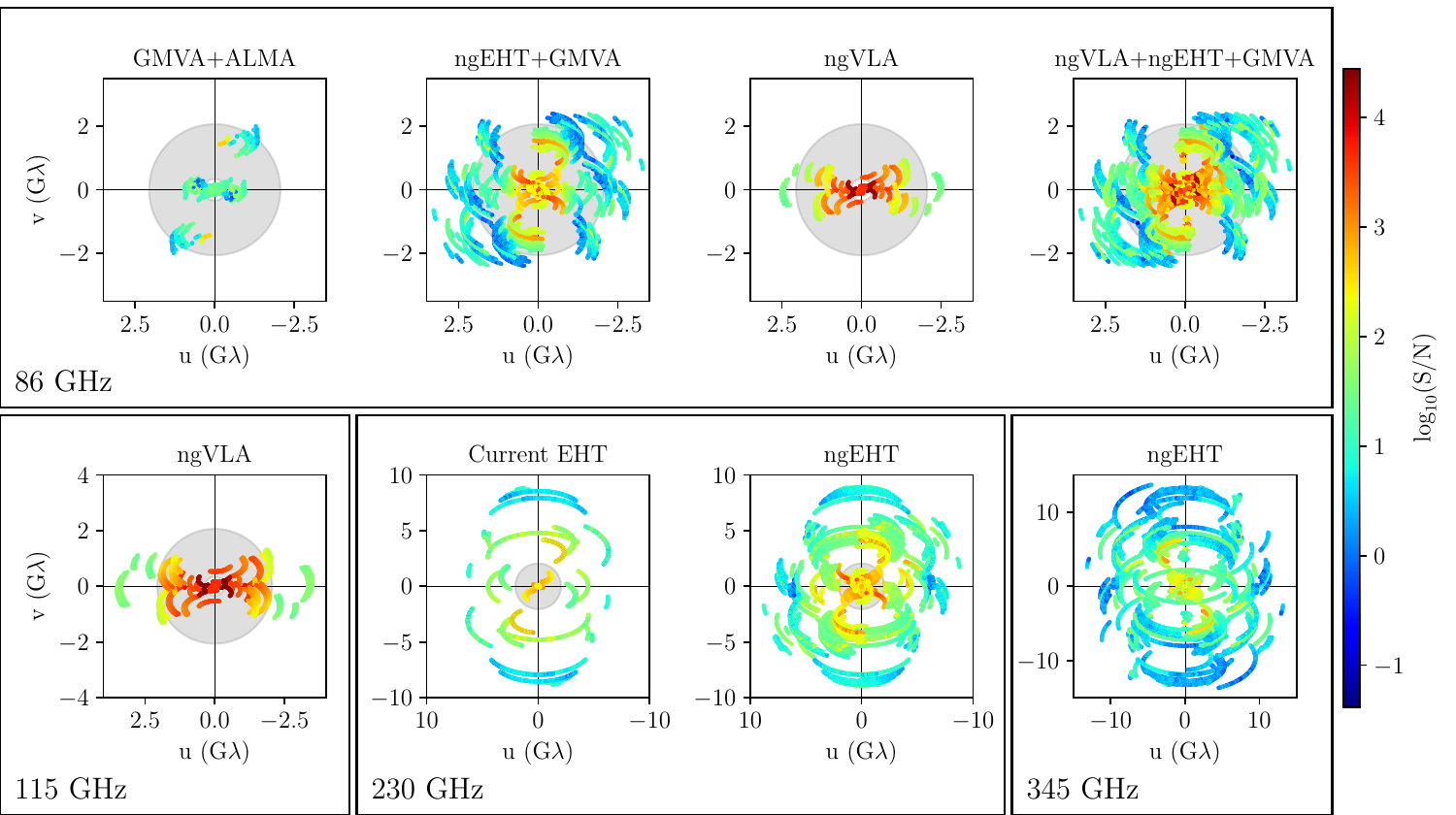}
    \caption{($u,v$) coverage plots for all arrays considered, with individual baselines colored by their signal-to-noise ratio ($\log_{10}(\textrm{S/N})$). The gray annuli delimit the region of baseline lengths which correspond to angular scales of 100 to 600\,$\mu$as on the sky: the region of ($u,v$) space most critical to jet detection. Arrays with the best prospects of detecting the diffuse, extended jet emission should have high S/N baselines in this shaded region.}
    \label{fig:uvcoveragearrays}
\end{figure*}

In this paper, we consider the prospects for detecting the jet in \sgra with a series of current and next-generation VLBI arrays.  
At 86\,GHz, we consider four different arrays: the currently operating GMVA+ALMA, the proposed ngVLA, a scenario where the ngEHT and GMVA are used in tandem, and a final scenario where the ngVLA, ngEHT, and GMVA are used as a single array.
To assess whether published non-detections of the jet are consistent with expectations from the simulation presented in Section \ref{sec:grmhd}, we have used the GMVA configuration that exactly correspond to the most recently published GMVA data on \sgra from \cite{issaoun2019}.
At 115\,GHz we only consider the ngVLA, as it uniquely offers 115\,GHz capabilities. At 230\,GHz, we consider the current EHT array and the proposed ngEHT array \citep {doeleman2023,roelofs2023}. At 345\,GHz we only considered the ngEHT as it provides the densest coverage for imaging \citep{roelofs2023}. Note that for the ngVLA we specifically consider the Long Baseline Array configuration, in which the core is phased into a single highly sensitive station and outer antennas are used as individual stations \citep[see, e.g.,][]{selina2018,issaoun2023}. \autoref{fig:uvcoveragearrays} shows the $(u,v)$ coverage plots and noise parameters for all the array and frequency combinations mentioned above.

\subsection{Synthetic Data Generation \label{sec:synthdata}}

To quantify which VLBI arrays are capable of detecting the jet, we perform image reconstruction tests that simulate observations of \sgra with the various VLBI arrays shown in Figure~\ref{fig:grmhd_snapshot_contours_grid}. We take the scattered GRMHD snapshots shown in the second row of Figure \ref{fig:grmhd_snapshot_contours_grid} as the reference ground truth images for \sgra at each frequency. We then generate synthetic complex visibilities (see Equation\,\ref{eq:measuredcomplexvis}), which simulate VLBI observations of these ground truth images with the various arrays.

These synthetic data were made using the Python package \texttt{ehtim} \citep{chael2023ehtim}. The array and noise parameters for GMVA at 86\,GHz were taken from existing 86\,GHz observations of \sgra from 2017 \citep{issaoun2019}. For all other arrays, we constructed a synthetic data set with idealized noise and antenna performance conditions. The synthetic observations start at 5 UTC and end at 16 UTC \citep[e.g.,][]{sgraIV}. Each scan lasts for five minutes, with a five-minute break in between scans. The integration time within each scan is 30 seconds. We also assumed a bandwidth of 2\,GHz, which is typical of stations in the EHT \citep{sgraIV}.

\begin{figure*}
    \centering
    \includegraphics[width=1.0
\textwidth]{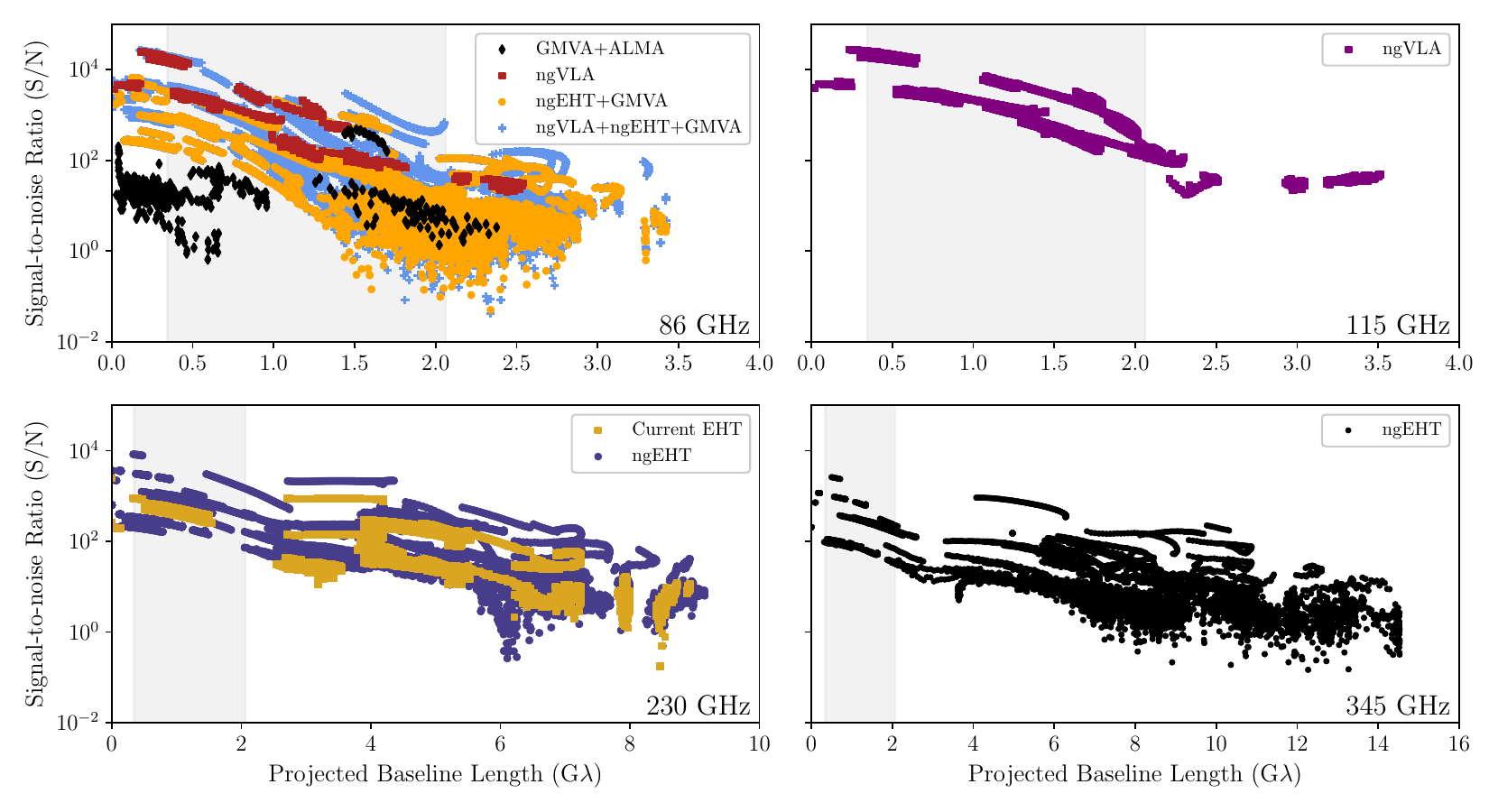}
    \caption{S/N as a function of projected baseline length for the arrays used to make image reconstructions in this paper. Top left: The 86\,GHz arrays considered: GMVA + ALMA \citep[black diamonds,][]{issaoun2019}, the ngVLA (\citep[red squares,][]{issaoun2023}), the ngEHT used in tandem with the GMVA \citep[orange circles; ngEHT parameters from][]{roelofs2023}, and the scenario in which ngVLA, ngEHT, and GMVA are all combined (blue crosses). Top Right: The 115\,GHz array considered, the ngVLA (purple squares). Bottom left: The 230\,GHz arrays considered: the current EHT array (yellow squares) and the ngEHT (purple circles) \citep{roelofs2023}. Bottom right: The 345\,GHz array considered: the ngEHT. All panels have a shaded region representing the range in projected baseline lengths that correspond to angular scales of 100\,$\mu$as to 600\,$\mu$as on the sky --- the scales most critical to jet detection.\label{fig:snr}}
\end{figure*}

\subsection{Image Reconstructions}

To reconstruct images of these synthetic data, we use \texttt{Comrade} (Composable Modeling of Radio Emission) \citep{tiede2022}, a fully Bayesian modeling package written in the \texttt{Julia} programming language \citep{bezanson2015julia}. This package is specifically designed for VLBI imaging of black holes and active galactic nuclei. Sparse sampling of image spatial scales with VLBI arrays leads to a wide range of possible images consistent with the data. By taking a fully Bayesian approach to fitting an image model to VLBI data, \texttt{Comrade} is able to model this uncertainty by exploring the posterior of possible images that fit the data. 

Here, we use \texttt{Comrade} to fit an image model to the synthetic complex visibilities generated in Section~\ref{sec:synthdata}\footnote{The code used in this section has been adapted from the \texttt{Comrade} Stokes I Simultaneous Image and Instrument Modeling Imaging tutorial found here: \url{https://ptiede.github.io/Comrade.jl/stable/tutorials/StokesIImaging}. This analysis was performed with Comrade v0.8.1.}. From Equation~\ref{eq:measuredcomplexvis}, the likelihood, or the probability that we measure the complex visibility $V_{ij}$ given an image model $I$, can be written as: 
\begin{equation}
    p(V_{ij}|I) = (2 \pi \sigma_{ij}^2)^{-1} \exp{\left(\frac{|V_{ij} - g_i g_j^* \, \mathcal{V}_{ij}|^2}{2 \sigma_{ij}^2} \right)}
    \label{eq:likelihood}
\end{equation}
where $V_{ij}$ is the measured visibility from the synthetic data, $\mathcal{V}_{ij}$ is the visibility predicted by our image model, $g_{ij}$ are instrument gains, and $\sigma_{ij}$ is the thermal noise. With this likelihood function, we can simultaneously fit the complex gains and the parameters describing our source image model.

\texttt{Comrade} supports image models which assume a source structure structure (e.g., a ring) and image models where no source structure is assumed (e.g., a grid of pixels). Our image model is a rasterized square grid of pixels, in which each pixel contains a variable amount of flux density. The pixel dimensions are chosen such that the pixel size spans $\frac{1}{4}$ to $\frac{1}{3}$ the angular resolution of the array being considered. The field of view of each image reconstruction was determined by the observing frequency: all 86 and 115\,GHz images were made with the full 600x600\,$\mu$as field of view of the ground truth images, while the 230\,GHz images were made with a 300x300\,$\mu$as and 150x150\,$\mu$as field of view. This was due to the sparser baseline coverage corresponding to large angular scales for the 230\,GHz arrays (see Figures~\ref{fig:uvcoveragearrays} and \ref{fig:snr}) resulting in difficulties in constraining image structure at larger scales.

The prior on our image model is a Gaussian Markov Random Field. We use a symmetric Gaussian distribution with a FWHM of $200\,\mu$as to represent the initial guess of the image structure and ensure smoothness. \texttt{Comrade} uses Hamiltonian Monte Carlo (HMC) to draw samples from the image posterior \citep{tiede2022}, where each sample corresponds to a different image. We begin by finding the maximum of the posterior, and we use this image as our starting point to initialize the HMC. The HMC is run for about 8000 steps, with each step corresponding to one sample of the posterior. The final images we show are obtained by averaging across the HMC samples and blurring to the observing array resolution with a circular Gaussian beam.

\section{Results}\label{sec:results}

\begin{figure*}
    \centering
    \includegraphics[width=1.0\textwidth]{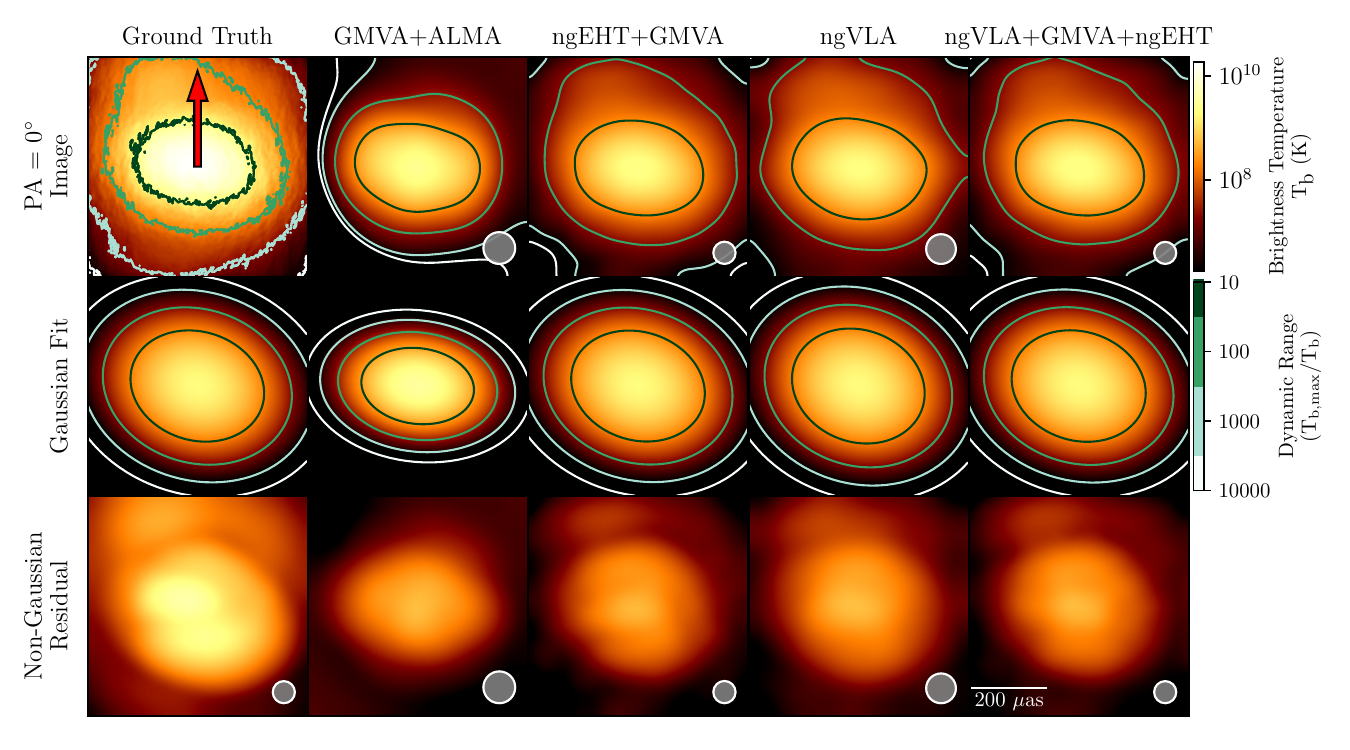}
    \caption{Top row: Image reconstructions of \sgra at 86\,GHz with \texttt{Comrade} using various array configurations. Here we consider the scenario where the jet is approximately perpendicular to the major axis of the scattering kernel, a red arrow shows the system position angle. This configuration explores detection prospects for the least challenging detection scenario: when the effects of interstellar scattering minimally obscure the jet. The array configurations and noise parameters are shown in Figures \ref{fig:uvcoveragearrays} and \ref{fig:snr}. Middle row: Two-dimensional Gaussian least-square fits to the corresponding images shown in the top row. Bottom row: The absolute value residual images produced from subtracting the corresponding top row and bottom row images. These residual images test if a non-Gaussian extended structure due to the jet is present in each image reconstruction.}
    \label{fig:86reconstructions}
\end{figure*}

\begin{figure*}
    \centering
    \includegraphics[width=1.0\textwidth]{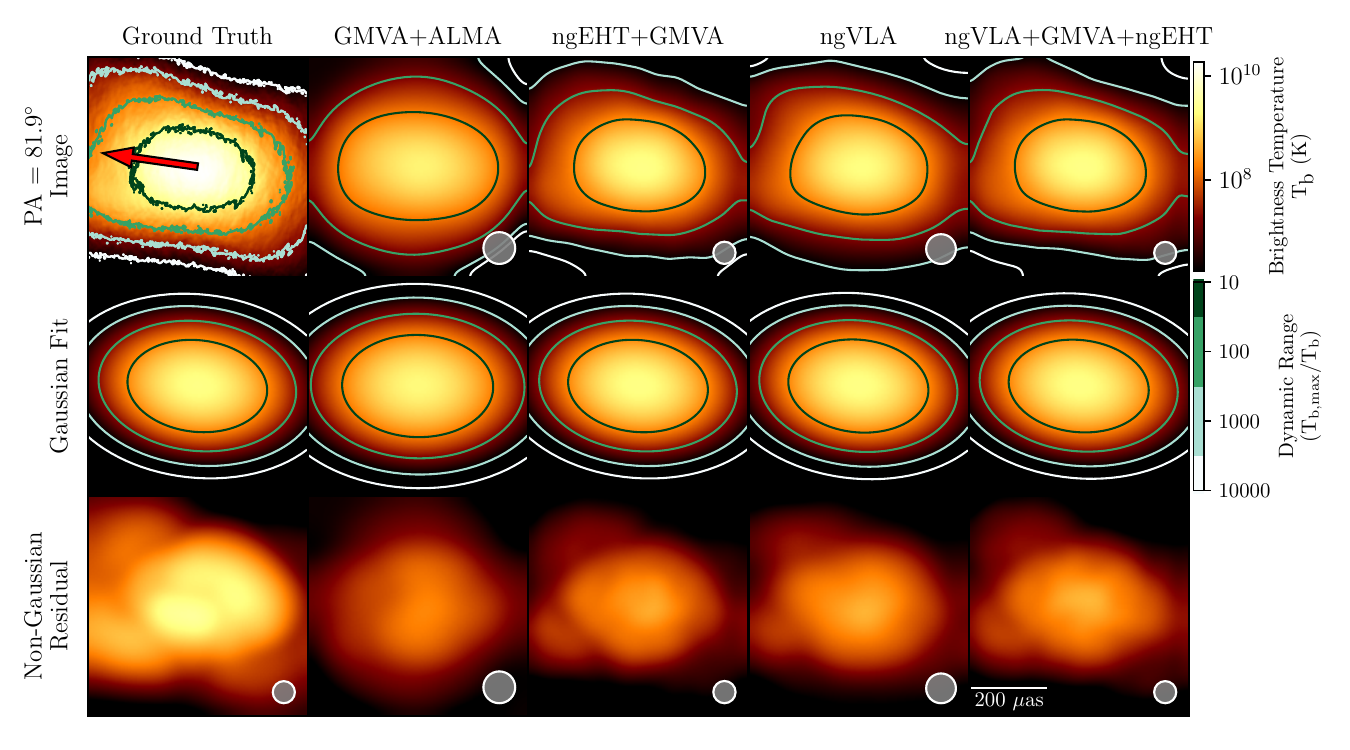}
    \caption{The same as Figure \ref{fig:86reconstructions}, now considering the scenario where the jet is aligned with the major axis of the scattering kernel. A red arrow shows the system position angle. This explores detection prospects for the most challenging detection scenario: when the effects of interstellar scattering maximally obscure the jet.}
    \label{fig:86reconstructionsaligned}
\end{figure*}

\begin{figure*}
    \centering
    \includegraphics[width=0.7\textwidth]{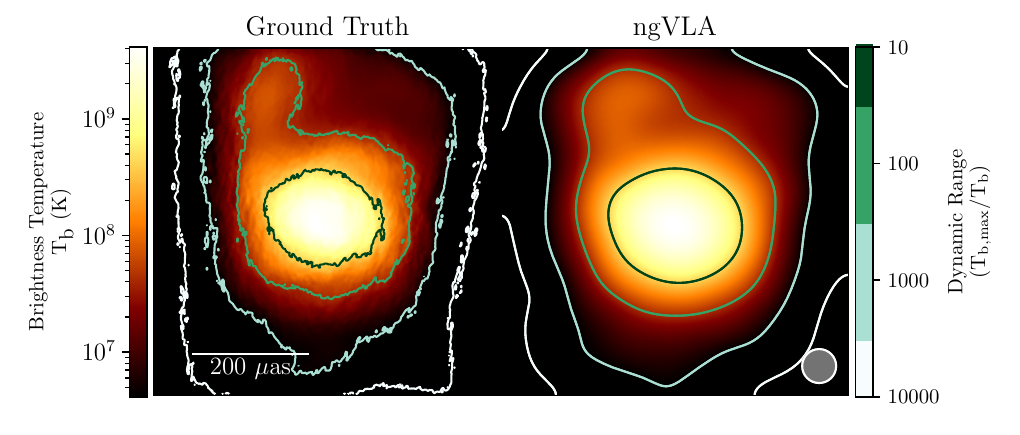}
    \caption{Image reconstruction of \sgra at 115\,GHz with the ngVLA made with \texttt{Comrade}. The image reconstruction has been blurred to the ngVLA's instrument resolution at 115\,GHz. Note that of the VLBI arrays we considered, only ngVLA has 115\,GHz capability. More intrinsic structure is recovered in comparison to 86\,GHz due to the relatively high jet power and a drastic reduction in interstellar scattering ($\theta_\text{scatt} \propto \nu^2$).}
    \label{fig:115reconstructions}
\end{figure*}

\begin{figure*}
    \centering
    \includegraphics[width=0.8\textwidth]{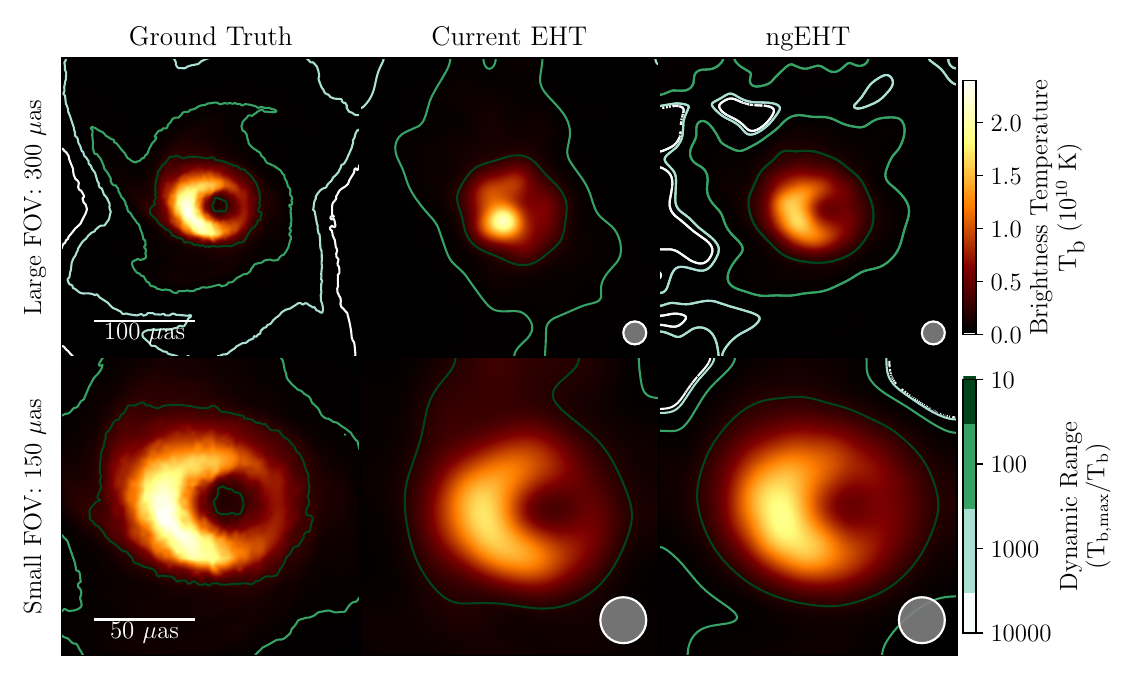}
    \caption{Image reconstructions of \sgra at 230\,GHz made with \texttt{Comrade} using the current EHT (middle column) and the ngEHT (rightmost column) array configurations. The array configurations and noise parameters are shown in Figures \ref{fig:uvcoveragearrays} and \ref{fig:snr}. The ground truth image (leftmost panels) is a scattered snapshot of the best-bet GRMHD model described in Section \ref{sec:grmhd}. Two imaging tests were done: one with an enlarged FOV ($300{\times}300 \, \mu$as, top row), and one with a small FOV ($150{\times}150 \, \mu$as, bottom row). }
    \label{fig:230reconstructioncomparison}
\end{figure*}

In this section, we present the results of our synthetic image reconstructions of \sgra for a series of VLBI arrays. The resulting image reconstructions at 86\,GHz, 115\,GHz, and 230\,GHz are shown in Figures~\ref{fig:86reconstructions}, \ref{fig:86reconstructionsaligned}, \ref{fig:115reconstructions}, and \ref{fig:230reconstructioncomparison}, respectively. 

First, we present our 86\,GHz imaging tests of \sgra with GMVA+ALMA, ngEHT+GMVA, ngVLA, and ngVLA+GMVA+ngEHT. The position angle of the black hole spin axis is unconstrained, hence the jet may be oriented along any direction on the sky (note that the inclination angle relative to the observer is constrained to $30\degree$, see Section~\ref{sec:grmhd}). The unconstrained position angle has implications for the impact of interstellar scattering on the image structure. Distortion of source structure due to interstellar scattering is most severe along the kernel major axis, which lies at a PA of $81.9 \degree$ (see \autoref{fig:grmhd_snapshot_contours_grid} and \autoref{sec:onskyimage_scattering}). This distortion is least severe along the kernel minor axis, which is generally in the north-south direction on the image. Therefore we performed two imaging tests at 86\,GHz: the first with a jet PA of $0 \degree$, and the second with a jet PA of $81.9 \degree$. The two tests approximate the ``best-case'' and ``worst-case'' scenarios for the jet orientation at 86\,GHz, in which the jet is minimally obscured and maximally obscured by interstellar scattering, respectively. The top rows of Figures \ref{fig:86reconstructions} and \ref{fig:86reconstructionsaligned} show our imaging results in these two limiting cases.

To quantify which image tests at 86\,GHz successfully capture the jet, we search for an image structure that is not associated with the central scatter-broadened region. We fit 2D Gaussian distributions to the ground truth image and each image reconstruction, and we subtract this Gaussian distribution from the corresponding image (see middle and bottom rows of Figures \ref{fig:86reconstructions} and \ref{fig:86reconstructionsaligned}). For an array to successfully image the jet, the corresponding image reconstruction must contain residual flux density outside the central region (where image structure is dominated by scatter broadening of the bright core) and this residual flux density must also be present in the ground truth image. From this metric, we determine that the currently operating GMVA+ALMA at 86\,GHz is unable to detect the predicted jet in both limiting cases. This is consistent with the latest observations of \sgra with this array configuration, which do not recover any extended emission at this frequency \citep{issaoun2019,issaoun2021}. However, all arrays that utilize next-generation VLBI (ngEHT+GMVA, ngVLA, and ngVLA+ngEHT+GMVA) successfully capture the jet for both jet orientations. The dynamic range of the image reconstructions is improved by about two orders of magnitude with the inclusion of any next-generation VLBI array.

In Figure~\ref{fig:115reconstructions} we present the results at 115\,GHz. Of the arrays we consider, only the ngVLA has the planned capability to observe this frequency. This image reconstruction quality is improved in comparison to 86\,GHz. The high dynamic range ($\gtrsim 1000$) and reduced effects of interstellar scattering result in an unambiguous jet detection, and the most intrinsic structure recovered in our imaging tests. 

At 230\,GHz we performed two separate imaging tests: one with the typical EHT FOV of $150 {\times} 150 \, \mu$as, and another with an enlarged FOV of $ 300 {\times} 300 \, \mu$as (see Figure \ref{fig:230reconstructioncomparison}). As expected, both the EHT and its upgrades through the ngEHT program successfully image the central ring of emission in both tests. However, neither array recovers the jet at this frequency. The EHT reconstructions have a dynamic range of $\sim 10$; any extended emission outside the central ring remains unconstrained. The ngEHT images display various improvements to the EHT images, they have a greater dynamic range of ${\sim}100$ and constrain extended structure on a larger angular scale, which successfully captures the inner accretion flow in both tests. However, capturing emission that is unambiguously associated with a jet at 230\,GHz requires a dynamic range of ${\sim}1000$, which is not achieved by either array. We did not generate image reconstructions at 345\,GHz, for which the prospects of jet detection are poorer because of more optically thin jet emission, worse baseline coverage, and worse sensitivity.

\section{Discussion}\label{sec:discussion}

Our imaging tests demonstrate that next-generation VLBI at 86 and 115\,GHz successfully capture the jet in optimal observing conditions, while VLBI at higher frequencies does not capture the jet. These findings are consistent with expectations from the images in \autoref{fig:grmhd_snapshot_contours_grid} and the associated discussion. We now discuss the specific properties of these VLBI arrays that influence jet recovery (\autoref{sec:array_requirements}) and the effects of simplifying assumptions made in our synthetic data generation and imaging (\autoref{sec:simplifications}). 

\subsection{Array Properties Needed for Jet Detection}
\label{sec:array_requirements}

As shown in the ground truth images (bottom row of Figure~\ref{fig:grmhd_snapshot_contours_grid}), the faint extended jet emission has structure on scales of approximately $100 \, \mu$as to $600 \, \mu$as. Therefore, VLBI arrays best suited to jet detection must contain high-sensitivity baselines that fill in the region of $(u,v)$ space that corresponds to these angular scales. Figures~\ref{fig:uvcoveragearrays} and \ref{fig:snr} show the baseline configurations and sensitivities of all the arrays we consider, the former in $(u,v)$ space and the latter as a function of projected baseline length. These plots both indicate the angular scales critical to jet detection mentioned before, which we use to evaluate the arrays. 

The VLBI arrays that failed to image the jet either have sparse coverage in this critical region of $(u,v)$ space (current EHT at 230\,GHz), or are not sensitive enough to capture the faint jet emission (GMVA+ALMA at 86\,GHz, ngEHT at 230\,GHz). While the ngEHT upgrades introduce new stations at scales relevant to jet detection, the sensitivity requirements to capture the faint jet emission are too severe at high radio frequencies. As mentioned above, the instrument sensitivity required for jet detection is an order-of-magnitude greater at 230\,GHz than at 86 and 115\,GHz. These issues are further compounded at 345\,GHz: the ngEHT baseline coverage at short baselines is sparser at 345\,GHz and the sensitivity requirements to capture jet emission are higher. As a result, we do not expect the ngEHT to successfully detect the jet in \sgra at 345\,GHz. 

Next-generation VLBI arrays at 86 and 115\,GHz offer more promising prospects for jet detection. At 86\,GHz, the ngEHT and ngVLA fill in critical gaps in $(u,v)$ space present in the GMVA+ALMA data. Note that we use the 2017 GMVA configuration in this analysis, but improvements in sensitivity and the addition of four stations in the following years \citep[e.g.,][]{Kim_2023} will further improve the prospects of jet detection at 86 GHz in future observations. The ngVLA also introduces the possibility of 115\,GHz observations (see \autoref{fig:grmhd_snapshot_contours_grid}), which is a ``sweet spot'' for jet detection. At this frequency the jet power is comparable to 86\,GHz, and the interstellar scattering is significantly reduced relative to 86\,GHz. In addition, the ngVLA is particularly suited for jet detection at 115\,GHz: the highest sensitivity ngVLA baselines lie in the critical region of $(u,v)$ space optimal for jet detection. As a result, we recovered more intrinsic jet structure in our ngVLA 115\,GHz reconstruction than at 86\,GHz. Future work may include implementing a scattering mitigation framework within \texttt{Comrade} to reduce the effects of scattering and recover more intrinsic source structure at 86 and 115 GHz \citep[for examples of scattering mitigation techniques see][]{johnson2016,issaoun2019}.

Note that these image tests were made for the GRMHD model described in Section \ref{sec:grmhd}. Here we briefly comment on how on the assumptions of this GRMHD model may impact jet detection prospects. Of the two best-bet models identified in \citetalias{sgraV} ($a_\star=0.5, 0.94$), the $a_\star=0.94$ simulation has an outflow power a factor of $\sim 3.5$ greater than $a_\star=0.5$ \citepalias[see Section 6 of][]{sgraV}. This is because these jets are powered by the extraction of black hole spin energy via the Blandford-Znajek mechanism \citep{blandfordznajek1977}, which results in low-spin black holes producing smaller, weaker jets. A smaller and dimmer jet may be more difficult to detect 86 GHz due to obscuration by interstellar scattering --- ngVLA observations at 115 GHz will be essential in this scenario. However, the inclusion of non-thermal electrons and emission with a magnetization parameter of $\sigma>1$ may result in a stronger, more extended jet due to additional emission from the jet spine \citep[e.g., Figure 12 of][]{Fromm2022}. Therefore our imaging tests may be a conservative estimate of how much extended jet structure is recoverable at $\sim 100$ GHz.

\begin{figure*}
    \centering
    \includegraphics[width=1\textwidth]{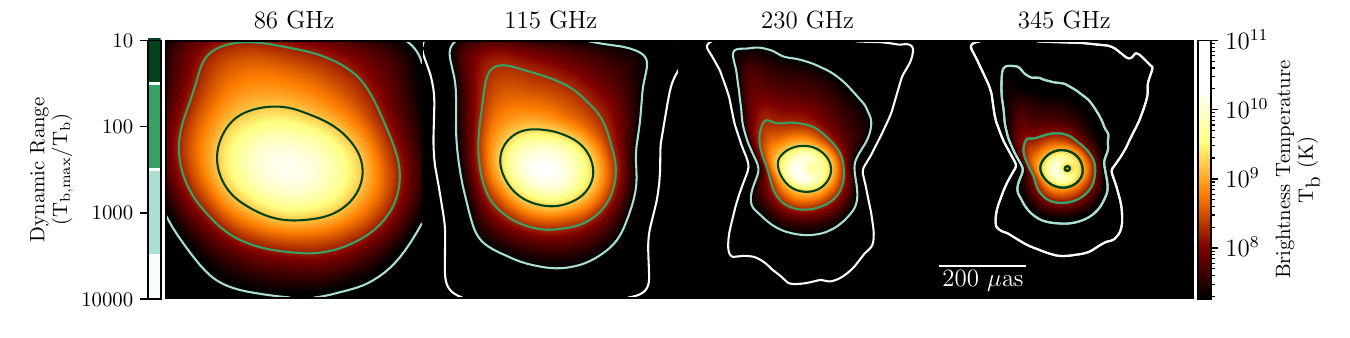}
    \caption{The time-averaged best-bet GRMHD simulation of \sgra shown at four different frequencies: 86, 115, 230, and 345\,GHz. The simulation length is 5 hours and 40 minutes. Each frame in the simulation is scattered with the same scattering realization and blurred to the resolution of an Earth-sized array. Logarithmic brightness contours (shown as ``dynamic range") are plotted to show the array sensitivity requirements to capture source image structure. Higher frequencies show greater variability, which in turn makes the prospects for jet detection more difficult.}
    \label{fig:averagesnapshots}
\end{figure*}

\subsection{Effects of Time Variability}
\label{sec:simplifications}

Throughout this analysis, we have made simplifications in generating our synthetic data and in our choice of source model. The most significant simplification is that we use a static GRMHD snapshot as our ground-truth image of \sgra. However, \sgra is variable on sub-hour timescales, and its variability presented a major challenge in producing the 2017 EHT imaging results \citep{sgraI}. Similarly, we expect variability to impact jet recovery.

To investigate the impact of variability on the jet structure, we take the full GRMHD simulation of \sgra ray-traced at the four frequencies of interest (86, 115, 230, and 345\,GHz), apply the same scattering realization to all the frames, and blur them to the resolution of an Earth-sized telescope. We then average these frames in time across the 5 hour and 40 minute simulation length (see Figure \ref{fig:averagesnapshots}). It is important to note that MAD GRMHD models --- including the model we use in this analysis --- over-predict the variability seen in \sgra \citep[see][]{sgraV,wielgus2022}. These time-averaged images quantify the extent of variability the predicted jet exhibits throughout a single observing run, and they represent an upper limit on the impact that variability \sgra may display during an actual observing run.

In Figure~\ref{fig:averagesnapshots}, we demonstrate that variability significantly affects the jet recovery via time-averaged images. Many of the sharp image features present in a single snapshot (see the bottom row of Figure \ref{fig:grmhd_snapshot_contours_grid}) are significantly broadened in the time-averaged images due to the source variability. In addition, the dynamic range contours in the time-averaged images are significantly less extended, and therefore variability further increases the sensitivity requirements for the VLBI arrays. The extent of this variability is also frequency dependent, with the higher frequency observations being significantly more impacted by source variability than lower frequency observations. The feature broadening is most significant at higher frequencies, where all sharp features have been completely averaged out, which in turn increases the instrument sensitivity requirements to capture extended structure. \sgra's source variability further complicates jet detection at 230 and 345\,GHz in particular, further reinforcing 86 and 115\,GHz as the most promising frequencies to detect the jet. 

Because of the extensive baseline coverage at 86 and 115\,GHz achieved by next-generation VLBI arrays, ``snapshot'' imaging should be possible and we expect that the prospects for jet detection will not be jeopardized by variability at these frequencies. This will be an important topic for detailed future study.

\section{Conclusions\label{sec:conclusions}}

\begin{figure*}
    \centering
    \includegraphics[width=1.0\textwidth]{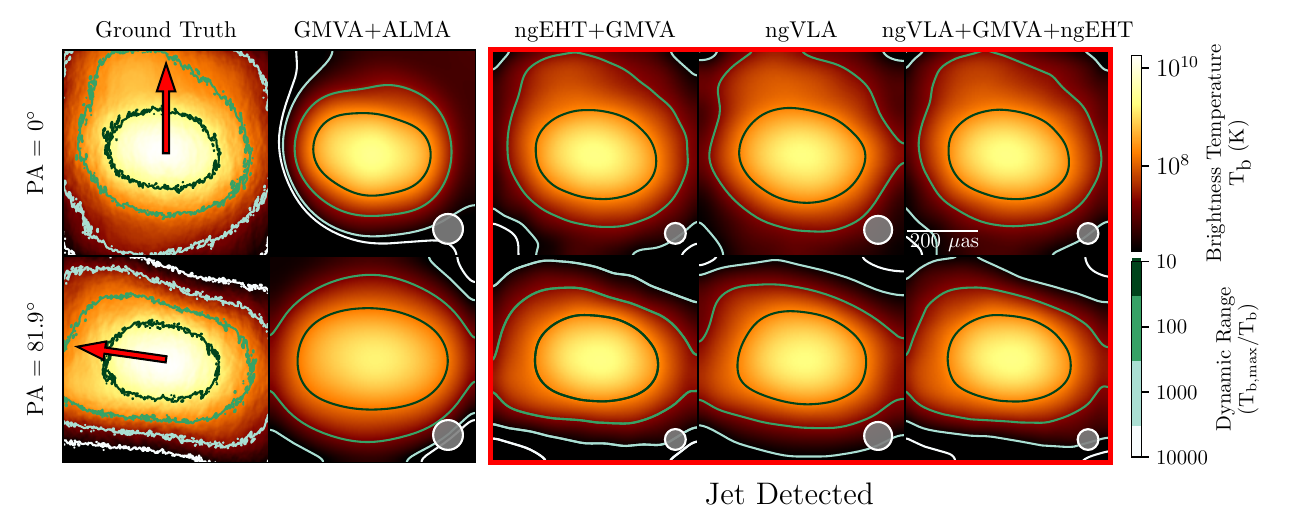}
    \caption{Image reconstructions of \sgra at 86\,GHz made with \texttt{Comrade} using the four array configurations we considered: GMVA+ALMA, ngEHT+GMVA, ngVLA, and ngVLA+GMVA+ngEHT. We have reconstructed images of two different jet orientations to represent the limiting cases regarding jet detection. The top and bottom rows show image reconstructions where the jet orientation is approximately perpendicular (PA=0$\degree$) and parallel (PA=81.9$\degree$) to the semi-major axis of the scattering kernel, respectively. Red arrows in both ground truth images point in the direction of the system PA. The image reconstructions that capture the jet are boxed in red: the ngEHT+GMVA, ngVLA, and ngVLA+GMVA+ngEHT are able to detect the jet in both limiting cases.}
    \label{fig:86reconstructioncomparison}
\end{figure*}

We have analyzed the prospects of jet detection in \sgra with current and future VLBI arrays. In particular, a combination of VLBI measurements and theory predict the existence of a jet in \sgra, but this jet has not been detected. This non-detection could either indicate inaccuracies in current simulations or limitations with current VLBI arrays.

Using the ``best-bet'' GRMHD simulation from the 2017 EHT campaign of \sgra \citep{sgraV,sgraVII}, we explored the prospects of detecting a jet with VLBI at four observing frequencies: 86, 115, 230, and 345\,GHz. This simulation produces a relativistic jet in \sgra. We have conducted imaging tests to determine if this predicted jet is consistent with non-detections in existing observations, and we have outlined an observing strategy with the highest prospects of jet detection. 

We have shown that interstellar scattering, intrinsic variability, and source opacity all give essential contributions to the image structure of \sgra and the prospects of jet detection at the different frequencies we consider. The $\nu^{-2}$ scaling of the interstellar scattering kernel size results in significant obscuration of the jet at 86\,GHz, but is almost negligible at 345\,GHz at the resolution of these VLBI arrays. As the observing frequency increases, the jet becomes more variable and optically thin. Generally, this makes the jet difficult to observe at frequencies above ${\sim}$100\,GHz. We have shown that images with a dynamic range of $\sim$100 are needed to capture extended jet emission at 86 and 115\,GHz, while a dynamic range of $\sim$1000 is needed at 230 and 345\,GHz. 

Our imaging tests demonstrated that the GMVA+ALMA at 86\,GHz, the current EHT at\,230 GHz, and the future ngEHT at\,230 GHz are unable to detect the predicted jet in \sgra, even with optimistic observing conditions. These arrays either lack the baselines needed to sample jet scales or lack the sensitivity needed to capture the faint jet emission. Hence, non-detections of the jet with currently available VLBI do not indicate problems in the simulations. Previously, \citet{Markoff_2007} showed that scattering can successfully hide the jet from VLBI observations at 43\,GHz.  

We show that the prospects for detecting the jet in \sgra with future arrays are promising, but primarily at ${\sim}100\,{\rm GHz}$. For instance, even after ngEHT upgrades, EHT observations at 345\,GHz are unlikely to detect the jet; the sensitivity requirements are more stringent, and it has sparser coverage on short baselines relative to 230\,GHz. However, our tests with next-generation VLBI arrays at 86 and 115\,GHz such as the ngEHT and the ngVLA do successfully detect the jet in GRMHD simulations. We attribute these successes to improvements in instrument sensitivity, particularly on baselines that sample scales relevant to jet detection (${\sim}100-600\,\mu{\rm as}$). Thus, 86\,GHz observations with upgraded arrays, and 115\,GHz observations that will be possible with the ngVLA, are the most promising paths to detecting a jet in \sgra.

While we have focused on detecting a jet through direct imaging, other pathways such as core-shift measurements using astrometry \citep[e.g.,][]{Moscibrodzka_2014,Fraga_2023,Jiang_2023} or chromatic time delays in light curves \citep[e.g.,][]{Brinkerink_2015,Brinkerink_2021} may also reveal the jet. While we have focused on total intensity measurements in this paper, the EHT has successfully resolved the polarization structure of \sgra on event-horizon scales \citep{sgraVII,sgraVIII}; polarization measurements may be another avenue to jet detection which can be explored in future investigations. Ultimately, the combination of these measurements will be crucial in revealing the nature of \sgra and its connection with the Galactic Center environment.

\begin{deluxetable*}{cccc}
\tablehead{
\colhead{Array} & \colhead{Frequency (GHz)} & \colhead{Operational Status} & \colhead{Jet Detection Status}
}
    \startdata
    GMVA+ALMA \citep{issaoun2019} & 86 & Online now & Not Detected \\
    EHT \citepalias{sgraI} & 230 & Online now & Not Detected \\
    ngEHT \citep{doeleman2023} & 230 & Online Late 2020s & Not Detected \\
    ngEHT+GMVA & 86 & Online Late 2020s & Detected \\
    ngVLA & 86 & Online 2035 & Detected \\
    ngVLA & 115 & Online 2035 & Detected \\
    ngVLA+ngEHT+GMVA & 86 & Online 2035 & Detected \\
    \enddata
    \caption{The results of the synthetic image reconstruction tests we performed. While the current VLBI is unable to capture the jet, next-generation VLBI arrays provide the improvements in baseline coverage and sensitivity needed to capture the jet.}
\end{deluxetable*}

\section*{Acknowledgements}
We would like to thank Maciek Wielgus for his constructive suggestions on this manuscript. This material is based upon work supported by the National Science Foundation Graduate Research Fellowship under Grant No. DGE 2140743. SI is supported by Hubble Fellowship grant HST-HF2-51482.001-A awarded by the Space Telescope Science Institute, which is operated by the Association of Universities for Research in Astronomy, Inc., for NASA, under contract NAS5-26555. 
We acknowledge financial support from the Brinson Foundation, the Gordon and Betty Moore Foundation (GBMF-10423), and the National Science Foundation (AST-2307887, AST-1935980, and AST-2034306). 
This work was supported by the Black Hole Initiative at Harvard University, which is funded by grants from the John Templeton Foundation and the Gordon and Betty Moore Foundation to Harvard University. CMF is supported by the DFG research grant ``Jet physics on horizon scales and beyond" (Grant No. 443220636).
YM is supported by the National Natural Science 
Foundation of China (grant No. 12273022), the Science and Technology Commission of Shanghai Municipality orientation program of basic research for international scientists (grant No. 22JC1410600), and the National Key R \& D Program of China (grant No. 2023YFE0101200).
The simulations were performed on LOEWE at the CSC-Frankfurt, Iboga at ITP Frankfurt, Siyuan Mark-I at Shanghai Jiao Tong University, and MISTRAL at the University of W\"urzburg.

\vspace{5mm}

\software{astropy \citep{astropy:2013, astropy:2018, astropy:2022},  
          Comrade \citep{tiede2022}, 
          ehtim \citep{chael2023ehtim},
          matplotlib \citep{Hunter2007matplotlib},
          numpy \citep{harris2020numpy},
          scipy \citep{2020SciPy-NMeth}
          pandas \citep{reback2020pandas}
          }

\appendix
\restartappendixnumbering
\section{Jet Power}

\begin{figure*}
    \centering
    \includegraphics[width=1\textwidth]{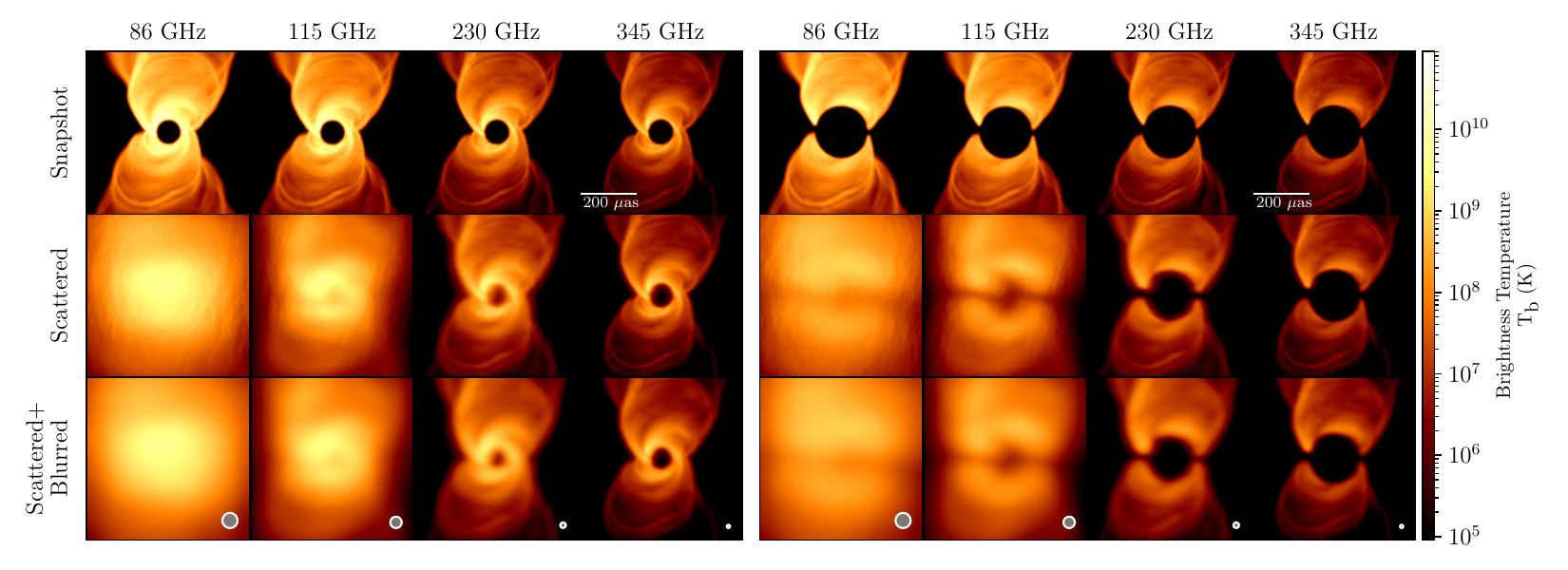}
    \caption{A grid showing the effects of scattering and blurring on the extended jet emission. A single GRMHD snapshot of \sgra is shown with a circular cut of diameter 100\,$\mu$as cut (left three columns) and 200 $\mu$as cut (right three columns) centered on the central black hole. These cuts are shown at 86, 115, 230, and 345\,GHz. The snapshots are then scattered (middle row) and blurred to the resolution of an earth-sized telescope at that frequency (bottom row). The images are scaled logarithmically to capture the diffuse extended emission. We repeat this process with the jet rotated at different position angles to produce Figure \ref{fig:tbmax_ratio}. While the effects of interstellar scattering and instrument resolution are minimized at higher frequencies, the extremely diffuse extended emission decreases jet detection prospects. \label{fig:100uas200uascutsgrid}}
\end{figure*}

To isolate the jet contribution to the flux density of \sgra in GRMHD snapshots, we eliminated all emission within a circular mask centered on the black hole. We repeated this for two circular masks, with diameters of $100\,\mu$as and $200\,\mu$as (see the top row of \autoref{fig:100uas200uascutsgrid}) to isolate the extended jet emission in each image. The resulting jet flux is shown in \autoref{tab:total_flux} --- the jet power decreases as the observing frequency increases from 86 to 345\,GHz as the emission becomes increasingly optically thin.

\begin{table}
\centering
\begin{tabular}{|l||cccc|}
 \hline
 & 86\,GHz & 115\,GHz & 230\,GHz & 345\,GHz \\
 \hline
\hline
 No Cut & 2.57 Jy & 2.7 Jy & 3.2 Jy & 3.1 Jy \\
 $100$ $\mu$as  Cut & 1.2 Jy & 1.0 Jy & 0.76 Jy & 0.57 Jy \\
 $200$ $\mu$as  Cut & 0.39 Jy & 0.29 Jy & 0.20 Jy & 0.14 Jy \\
 \hline
\end{tabular}
\caption{Total time-averaged flux density of the GRMHD simulation of \sgra at the chosen observing frequencies. The 100 and 200\,$\mu$as cuts refer to the total flux when the central region of the image is masked out with a circular mask of 100 or 200\,$\mu$as, respectively, to isolate the flux contribution from the jet. Note that the jet power decreases with frequency.\label{tab:total_flux}}
\end{table}

\begin{figure*}
    \centering
    \includegraphics[width=1\textwidth]{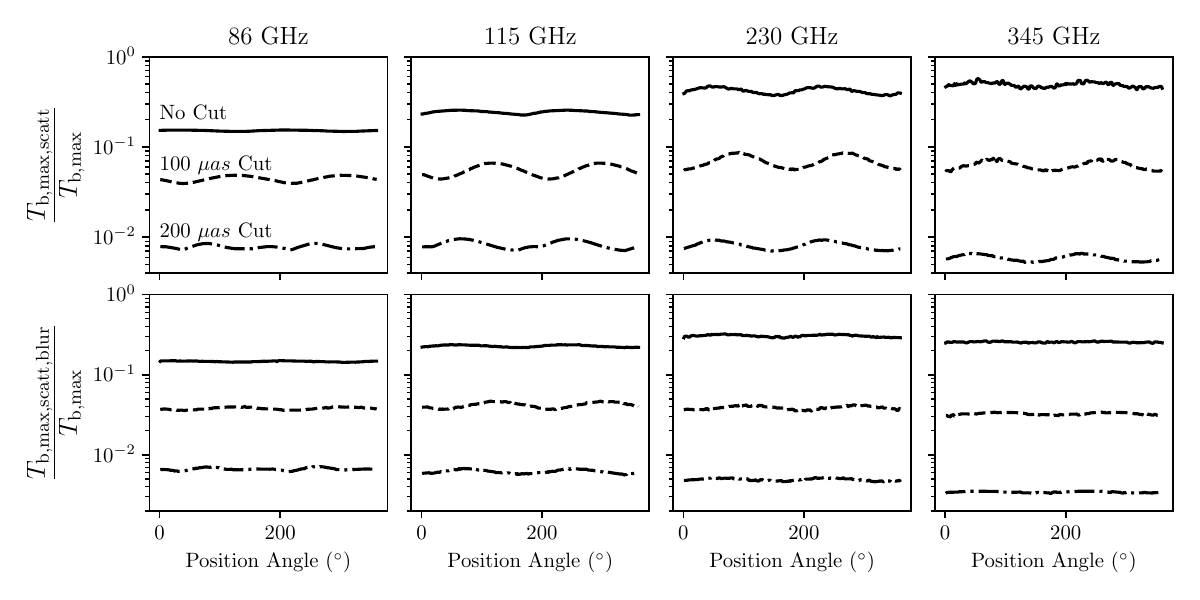}
    \caption{Top row: The ratio of the maximum brightness temperature in a scattered image versus the corresponding unscattered image $(T_{\mathrm{b,max, scatt}}/T_{\mathrm{b,max, unscatt}})$ measured as a function of position angle (PA) at 86, 115, 230, and 345\,GHz. This ratio was measured for three cases: 1. the original snapshot 2. the snapshot with a $100 ~\mu as$ diameter circular cut centered on the black hole 3. A the snapshot with a $200 ~\mu as$ diameter circular cut centered on the black hole.
    Bottom: The ratio of the maximum brightness temperature in a scattered and blurred image versus the corresponding unscattered and unblurred image $(T_{\mathrm{b,max, scatt,blur}}/T_{\mathrm{b,max, unscatt, unblur}})$ measured as a function of position angle (PA) at the four frequencies. This is repeated for the same three cases shown in the top panels. These plots characterize the flux loss a GRMHD snapshot due to interstellar scattering and by blurring the images to resolution. The 100 and 200 $\mu$as cuts show how the flux loss affects emission from the jet.}
    \label{fig:tbmax_ratio}
\end{figure*}

\bibliography{references}{}
\bibliographystyle{aasjournal}

\end{document}